\documentclass[%
reprint,pra,
superscriptaddress,
amsmath,amssymb,
aps,
]{revtex4}

\usepackage{graphicx}% Include figure files
\usepackage{dcolumn}% Align table columns on decimal point
\usepackage{bm}% bold math
\usepackage{hyperref}% add hypertext capabilities
\usepackage[utf8]{inputenc}
\usepackage{subfigure}
\usepackage{cleveref}
\usepackage{amsmath}
\begin{document}
	
\preprint{APS/123-QED}

\title{Sub-Nyquist computational ghost imaging with orthonormalized colored noise speckle patterns}

\author{Xiaoyu Nie}
\affiliation{%
	Texas A\&M University, College Station, Texas, 77843, USA}%
\affiliation{%
	Xi'an Jiaotong University, Xi'an, Shaanxi 710049, China}%
\author{Xingchen Zhao}%
\affiliation{%
	Texas A\&M University, College Station, Texas, 77843, USA}%
\author{Tao Peng}%
\email{taopeng@tamu.edu}
\affiliation{%
	Texas A\&M University, College Station, Texas, 77843, USA}%
\author{Marlan O. Scully}%
\affiliation{%
	Texas A\&M University, College Station, Texas, 77843, USA}%
\affiliation{%
Baylor University, Waco, 76706, USA}%
\affiliation{%
Princeton University, Princeton, NJ 08544, USA}%
\date{\today}

\begin{abstract}
Computational ghost imaging generally requires a large number of patterns to obtain a high-quality image. Both pre-modulated orthogonal patterns and post-processing orthonormalization methods have been demonstrated to reduce the pattern number and increase the imaging quality. In this work, We propose and experimentally demonstrate an orthonormalization approach based on the colored noise speckle patterns to achieve sub-Nyquist computational ghost imaging. We exam the reconstructed image in quality indicators such as the contrast-to-noise ratio, the mean square error, the peak signal to noise ratio, and the correlation coefficient. The results suggest that our method can provide a high-quality image using a sampling ratio an order lower than the conventional methods. The results also suggest that there exist an optimal frame rate under the noise environment.
\end{abstract}

	\maketitle
	
Computational ghost imaging (CGI) \cite{shapiro2008computational,bromberg2009ghost}, an ameliorated scheme on traditional ghost imaging (GI) \cite{Pittman1995Optical,bennink2004quantum,zhang2005correlated}, owns the ability to reconstruct the object via a single-pixel detector. CGI also grants advantages in an expanding range of non-conventional applications such as wide spectrum imaging \cite{edgar2015simultaneous,radwell2014single} and depth mapping \cite{howland2013photon,sun2016single}. It also finds application to various fields, such as temporal imaging \cite{devaux2016computational}, X-ray imaging \cite{klein2019x}, and remote sensing \cite{erkmen2012computational}. However, its sampling number is usually comparable to the total number of pixels in the speckle pattern to ensure good imaging quality, thus is time-consuming and resource-intensive. It also produces limits such as only suitable for static object reconstruction. 

Various methods have been proposed to overcome this problem.
Compressing sensing is a well-known technique that
reduces the required sample rate to lower than 30\% by exploring
the properties of sparsity  ~\cite{katz2009compressive,katkovnik2012compressive}. Still, it is also strictly limited
by the sparsity of the image. Deep learning has also shown its
ability to achieve sub-Nyquist imaging ~\cite{lyu2017deep,he2018ghost,wu2020sub}. The limitation is that most of the networks are trained by experimental CGI results. Therefore numerous measurements have to been done in advance. In addition, the environment and the training inputs for image reconstruction have to be made almost identical to the training environment and similar to the tested objects to make the system effective. Another approach is to use the orthonormalized pattern to reduce the sample rate~\cite{sun2017russian, luo2018orthonormalization}. In particular, Luo \textit{et al.} introduced a data post-processing algorithm to improve the reconstructing process in a GI system with pseudo-thermal light~\cite{luo2018orthonormalization}. The required sampling number is reduced by applying the Gram-Schmidt process to the noise patterns and intensity sequence collected by the bucket detector. However, such a method is sensitive to noise, and the image quality is not comparable with standard CGI when the sampling rate is high. Traditionally, Gaussian white noise speckle pattern is used for GI. We recently developed a method to generate the colored noise speckle pattern for CGI by customizing the speckle patterns' power spectrum distribution~\cite{li2021sub,nie2020noise}. Unlike white noise, colored noise generally has non-zero cross-correlation between neighborhood pixels. Sub-Rayleigh imaging was demonstrated with the blue noise pattern, which has negative cross-correlation between two adjacent pixels~\cite{li2021sub}. The pink noise pattern allowed us to image in a variety of noisy environments~\cite{nie2020noise}. 
	
In this letter, we present an orthonormalization method that combines the colored noise technique, thereby reducing the sampling in the CGI experiment significantly. We also compare the orthonormalization colored noise GI (OCGI) with orthonormalization white noise GI (OWGI), traditional white noise GI (WGI), and pink noise GI (PGI). The results are tested using the quality indicators such as the Contrast-to-Noise Ratio (CNR), the Peak Signal to Noise Ratio (PSNR), the Correlation Coefficient (CC), and the Mean Square Error (MSE). We show that OCGI always has the best performance. It can reduce the sample rate one order lower while still obtaining the same image quality as standard CGI. In addition, it suggests an optimal frame rate ($<1$) in the presence of noise. 
	
\begin{figure}[h!]
		\centering\includegraphics[width=0.95\linewidth]{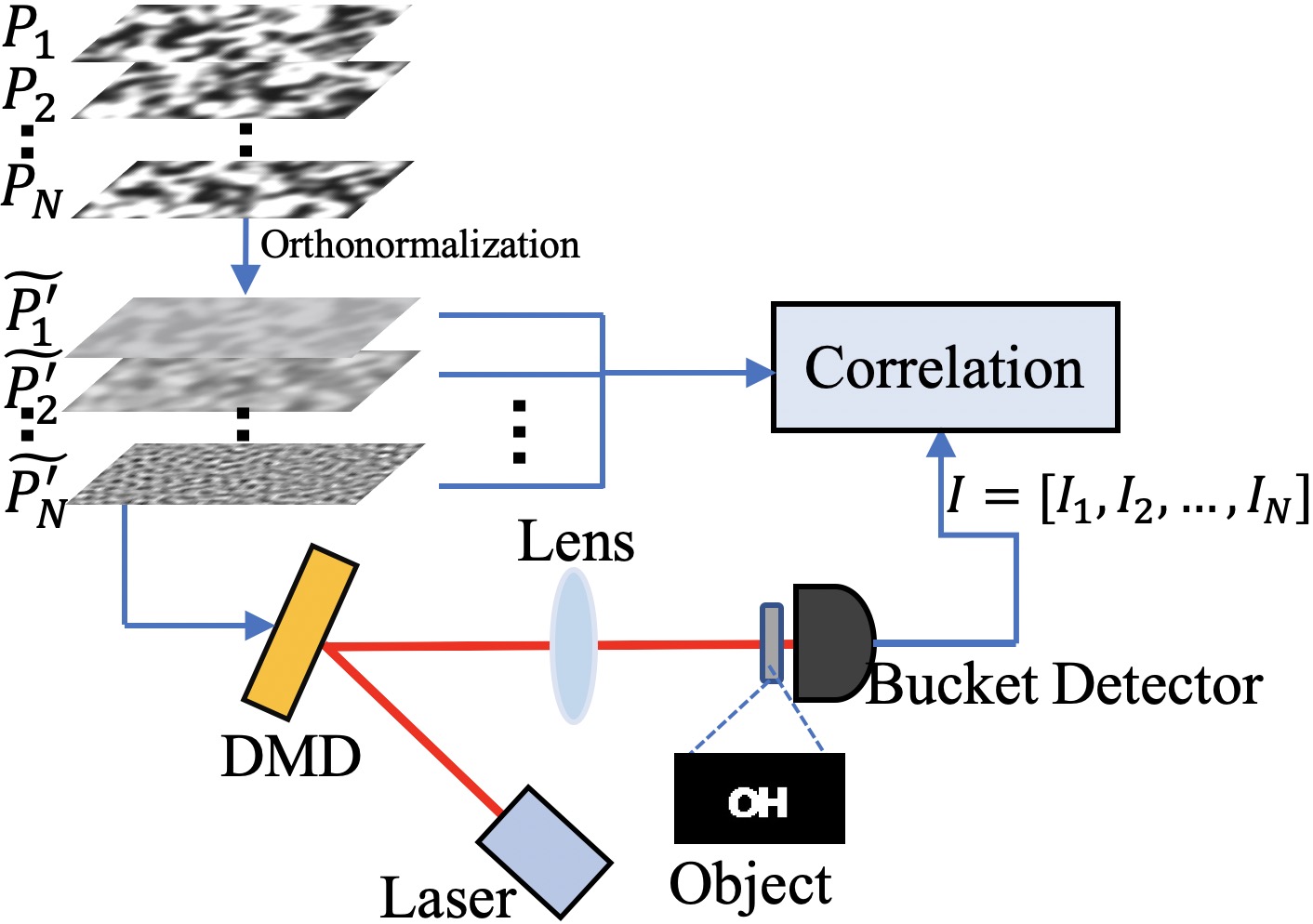}
		\caption{Schematic of the setup. The digital micromirror device (DMD) is illuminated by a CW laser. Orthonormalized patterns are loaded on the DMD then imaged onto the object plane. Correlation measurement is made between the patterns and the intensities recorded by the bucket detector.}
		\label{fig:setup}
	\end{figure}

The experimental setup is shown in Fig.~\ref{fig:setup}. This is a typical CGI setup: a CW laser illuminates the digital micromirror device (DMD), where the speckle patterns with designed distributions are loaded. The pattern generated by the DMD is then projected onto the object with the letters `OH' etched on an opaque plate. A bucket detector is put right after the object to record the transmitted light intensity. The DMD contains tiny pixel-mirrors each measuring $16\mu m \times 16\mu m$. Each noise pattern has $54\times 98$ independent pixels in the experiment, and each independent pixel consists of $10 \times 10$ mirrors. 
	
\begin{figure}[htbp]
		\centering\includegraphics[width=0.95\linewidth]{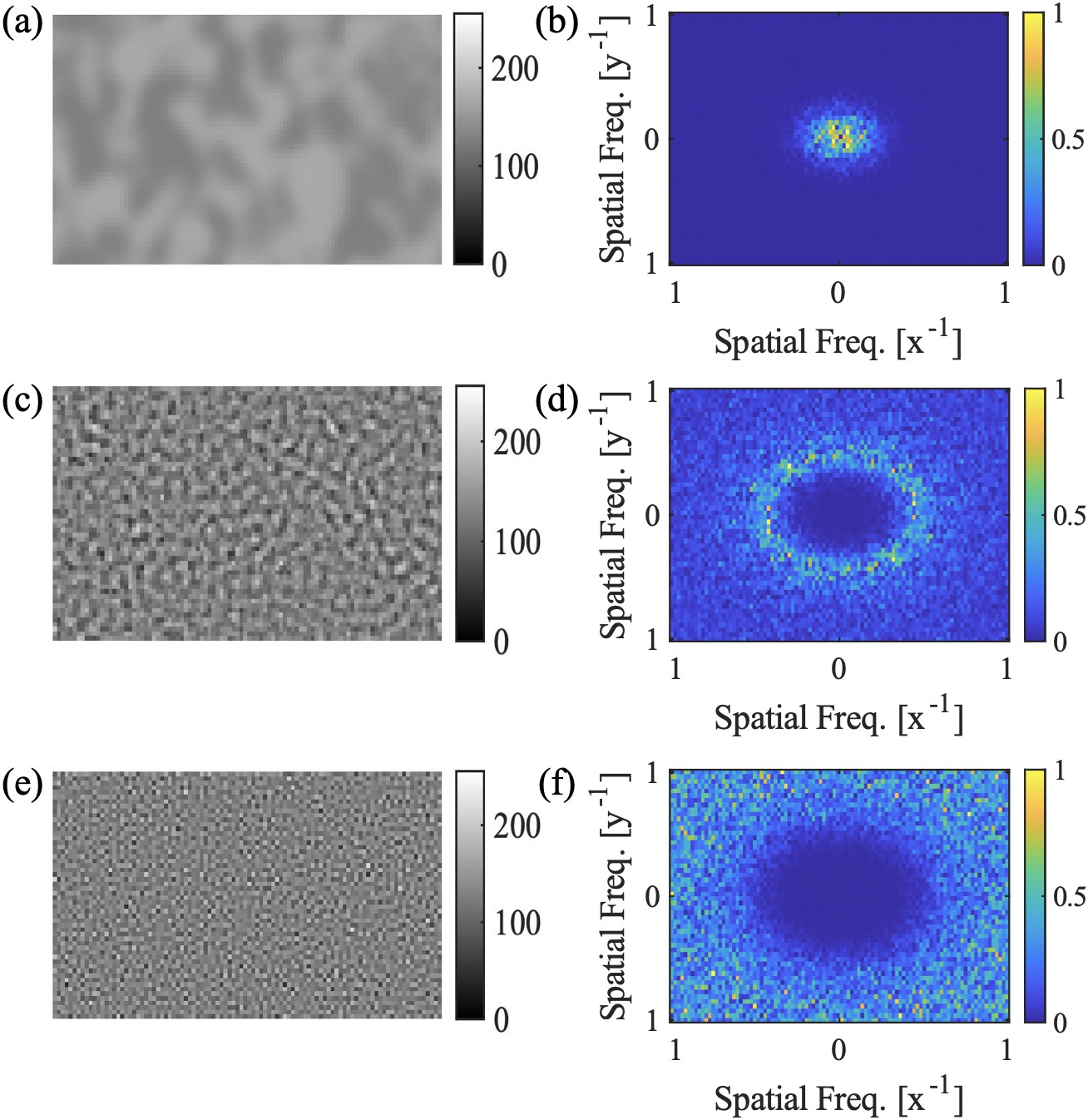}  
		\caption{The orthonormalized colored noise pattern: (a) the 1st pattern, (c) the 1000th pattern, (e) the last pattern (5292th); (b), (d), and (f) are normalized spatial frequency distributions of the 1st pattern, the 1000th pattern, and the 5292th pattern, respectively.}
		\label{fig:noisefeature}
	\end{figure}

	Firstly, the Gaussian white and pink noise patterns are generated by applying inverse Fourier transformation upon the spectrum in which the spatial frequencies are defined as ${\omega ^0}$ and ${\omega ^{-1}}$ \cite{nie2020noise}. Random phase matrices are also assigned to each pattern. The Gram-Schmidt process is then performed to orthonormalize the patterns. After the orthonormalization, the pink noise pattern's spatial frequency gradually changes to blue noise distribution, as shown in Fig.~\ref{fig:noisefeature}. In other words, the spatial frequency of the orthonormalized pattern covers a broad spatial spectrum range from pink to white and blue. The initial colored patterns are matrices $P_1, P_2, P_3, \cdots, P_{\mathrm{N}}$, and the orthonormalized patterns are matrices $\widetilde{P}_1, \widetilde{P}_2, \widetilde{P}_3, \cdots, \widetilde{P}_{\mathrm{N}}$, all of which contain $54 \times 98$ elements. We define the projection coefficient $c_{\mathrm{mn}}$ as
	\begin{equation}
	c_{\mathrm{mn}}=\frac{{P_\mathrm{m}}\cdot{\widetilde{P}_{\mathrm{n}}}}{\widetilde{P}_{\mathrm{n}}\cdot{\widetilde{P}_{\mathrm{n}}}}.
	\end{equation}
	The orthonormalized patterns can be generated by
	\begin{equation}
	\widetilde{P}_1=P_1,
	\end{equation}
	\begin{equation}
	\widetilde{P}_{\mathrm{m}}=P_{\mathrm{m}}-\sum\limits_{n=1}^{m-1}c_{\mathrm{mn}}\widetilde{P}_{\mathrm{n}}.
	\end{equation}
	Then, we re-normalize the histogram of $\widetilde{P}_1, \widetilde{P}_2, \widetilde{P}_3, \cdots, \widetilde{P}_{_{\mathrm{N}}}$ to [0, 255], which we define as $\widetilde{P}'_1, \widetilde{P}'_2, \widetilde{P}'_3, ......, \widetilde{P}'_{\mathrm{N}}$. According to the number of orthogonal vector space, we generate 5292 patterns for each kind, which is equal to the number of total pixel in a single pattern. We note here that, unlike the post-processing method shown in~\cite{luo2018orthonormalization}, we directly generate these orthonormalized patterns and apply them to DMD. Therefore, the orthonormalization coefficients and patterns are made at once. Besides, we don't have any intensity losses during the orthonormalization process. In our scheme, the intensity is measured as 
	\begin{equation}
	I_{\mathrm{i}}=T\cdot{\widetilde{P}'_{\mathrm{i}}},
	\end{equation}
	where T is the object's transmission coefficient, $\widetilde{P}'_{\mathrm{i}}$ is the i-$\mathrm{th}$ orthonormalized pattern. As shown in Fig.~\ref{fig:setup}, the image is retrieved by calculating the correlation between patterns and collected light intensity sequence as
	\begin{equation}
	\Gamma^{(2)}=\frac{1}{N}\sum\limits_{i=1}^{N}{\widetilde{P}'_{\mathrm{i}}{I_{\mathrm{i}}}}-\frac{1}{N^2}\sum\limits_{i=1}^{N}{\widetilde{P}'_{\mathrm{i}}}\times\sum\limits_{i=1}^{N}{{I_{\mathrm{i}}}},
	\end{equation}
	where ${N}$ is the sampling number. We define $\beta$ as the sampling rate, which is the ratio between the sampling number $N$ and the number of speckle in each pattern $N_{\mathrm{pixel}}$:
	\begin{equation}
	{\beta} = \frac{N}{N_{\mathrm{pixel}}}.
	\end{equation}
	
\begin{figure}[htbp]
		\centering\includegraphics[width=\linewidth]{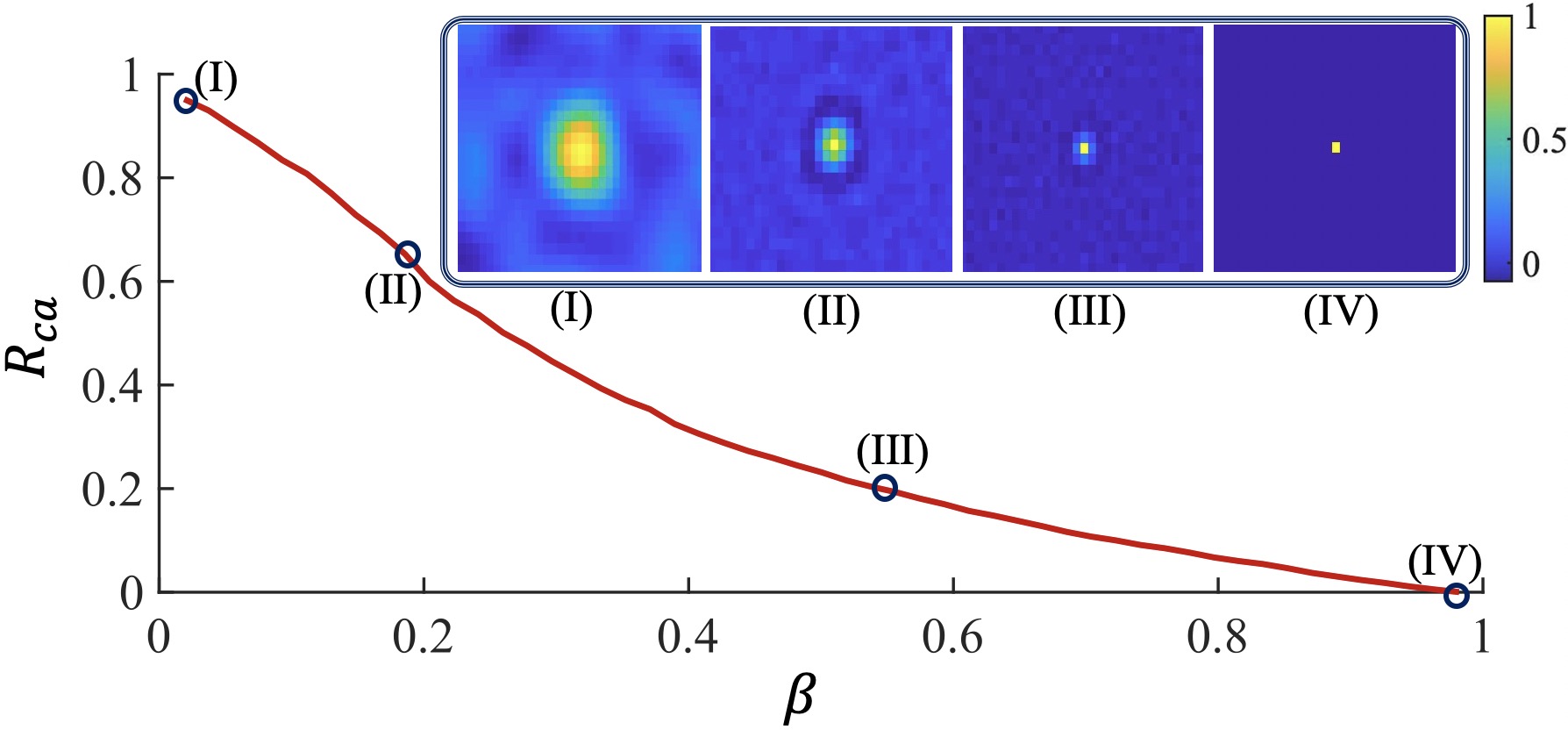}
		\caption{Cross-auto correlation ratio $R_{ca}$ as a function of the sampling rate $\beta$. Inserted pictures: (I), (II), (III), and (IV) are 2D plotted auto- and cross-correlation of total pattern number 100, 1000, 3000, and 5292, respectively.}
		\label{fig:correlation}
	\end{figure}

	We explore the properties of the orthonormalized pattern by analyzing its spatial frequency, auto-, and cross-correlation. As shown in Fig.~\ref{fig:noisefeature}, when the pattern number increases, the frequency peak moves to the higher end. This suggests that the pattern gradually changes from pink noise distribution to blue noise distribution under the orthonormalization process. This is easy to understand since the orthonormalization protocol naturally requires that the spatial frequency domain is also under orhtnormalization. Therefore, the OCGI maintains the pink noise's advantage when the sampling number is small, and it can continuously enhance the resolution when increasing the sampling number. Indeed, the OCGI owns the OWGI's feature when $\beta$ approaches 1, as shown in Fig.~\ref{fig:correlation}. 
		A random pixel ${p(x,y)}$ is chosen and its auto-correlation and cross-correlation with all other pixels are calculated. The cross-auto correlation ratio $R_{{\mathrm{ca}}}$ is defined as,
	\begin{equation}
	R_{{\mathrm{ca}}} = \frac{\Gamma^{(2)}_{p(x-1,y)}+\Gamma^{(2)}_{p(x+1,y)}+\Gamma^{(2)}_{p(x,y-1)}+\Gamma^{(2)}_{p(x,y+1)}}{4\Gamma^{(2)}_{p(x,y)}}.
	\end{equation}
	 From the pink line in Fig.~\ref{fig:correlation}, we can see that the ratio is gradually dwindling. the cross-correlation starts from nearly 1 when $\beta$ is small.  It then gradually decreases to 0 when $\beta=1$, which is the same as the white noise pattern. In a matter of fact, from the spatial frequency distribution of an arbitrary pattern, we can precisely predict the change of result during the image retrieving process with the OCGI method. It is also expected that the OCGI and OWGI measurements will converge to the same results when $\beta$ approaches 1, as shown in the following. 
	
	%\section{Simulation Results}
	\begin{figure}[htbp]
		\centering\includegraphics[width=0.95\linewidth]{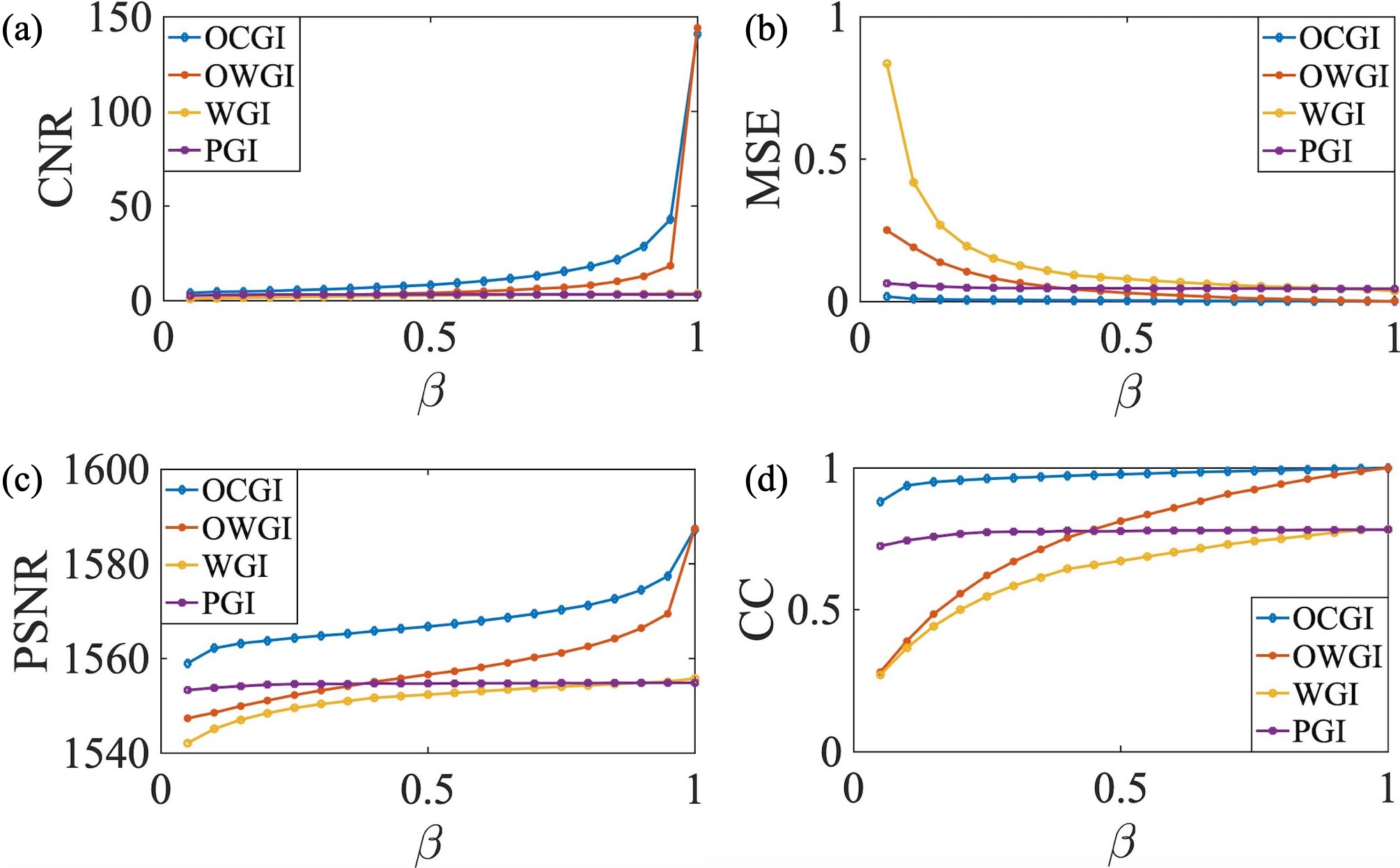}
		\caption{Simulation with no noise. Image qualities via different sampling numbers by CGI in the ideal condition. (a) CNR, (b) MSE, (c) PSNR, and (d) CC. }
		\label{fig:simu_NOnoise}
	\end{figure}
	
	To test the feasibility of the OCGI method, we run a simulation firstly in the ideal condition without any noise. To better judge the performance of various methods, \textit{i.e.}, the OCGI, OWGI, WGI, and PGI, we utilize four evaluating indicators of image quality, \textit{i.e.}, CNR, MSE, PSNR, and CC \cite{zerom2012thermal,xu2015optimization,li2017image,luo2018orthonormalization}:
	\begin{equation}
	{CNR} = \frac{{\langle G_{(\mathrm{o})}\rangle -\langle G_{(\mathrm{b})}\rangle }}{\sqrt{Var[G_{(\mathrm{o})}]+Var[G_{(\mathrm{b})}]}}
	\end{equation}
	\begin{equation}
	%{MSE} = \frac{1}{N_{pixel}}\sum{\frac{(G_{(1)}-{<G_{(1)}>})^2+{(G_{(0)}-{<G_{(0)}>})^2}}{<G_{(1)}>^2}}
	{MSE} = \frac{1}{N_{\mathrm{pixel}}}\sum_{i=1}^{N_{\mathrm{pixel}}}{[\frac{G_\mathrm{i}-X_\mathrm{o}}{\langle G_{(\mathrm{o})}\rangle}]^2}
	\end{equation}
	\begin{equation}
	{PSNR} = 10\times{log_{10}[\frac{(2^k-1)^2}{MSE}]}
	\end{equation}
	\begin{equation}
	{CC} = \frac{Cov(G,X)}{\sqrt{Var(G)Var(X)}}
	\end{equation}
	Here, $X$ is the reference matrix calculated by
	\begin{equation}
	{X_i} =  
	\begin{cases}
	\langle G_{(\mathrm{o})}\rangle & \text{, Transmission = 1}\\
	\langle G_{(\mathrm{b})}\rangle & \text{, Transmission = 0}
	\end{cases}
	\end{equation}
	$G_{(\mathrm{o})}$ represents pixels in the correlation results that the light ought to be transmitted, \textit{i.e.}, the object area, while $G_{(\mathrm{b})}$ represents pixels in the correlation results that the light ought to be blocked, \textit{i.e.}, the background area. $k$ is the gray level of the image, and in our experiment $k\equiv 8$. 
	
	As shown in Fig.~\ref{fig:simu_NOnoise}, the OCGI, similar to the PGI, gives a stronger signal in the low sample rate domain, as demonstrated in our previous study\cite{nie2020noise}. Besides, the OCGI always has the best image quality. The OWGI and OCGI have almost the same behavior when the image quality is saturated as the pattern number reaches the maximum. Both of them are still much better than WGI and PGI. Here the orthonormalization process completely smears the weakness of PGI, whose image quality is almost smooth from the beginning to the end, and strengthens the PGI's advantage at the low sampling level.
	
\begin{figure}[htbp]
		\centering\includegraphics[width=\linewidth]{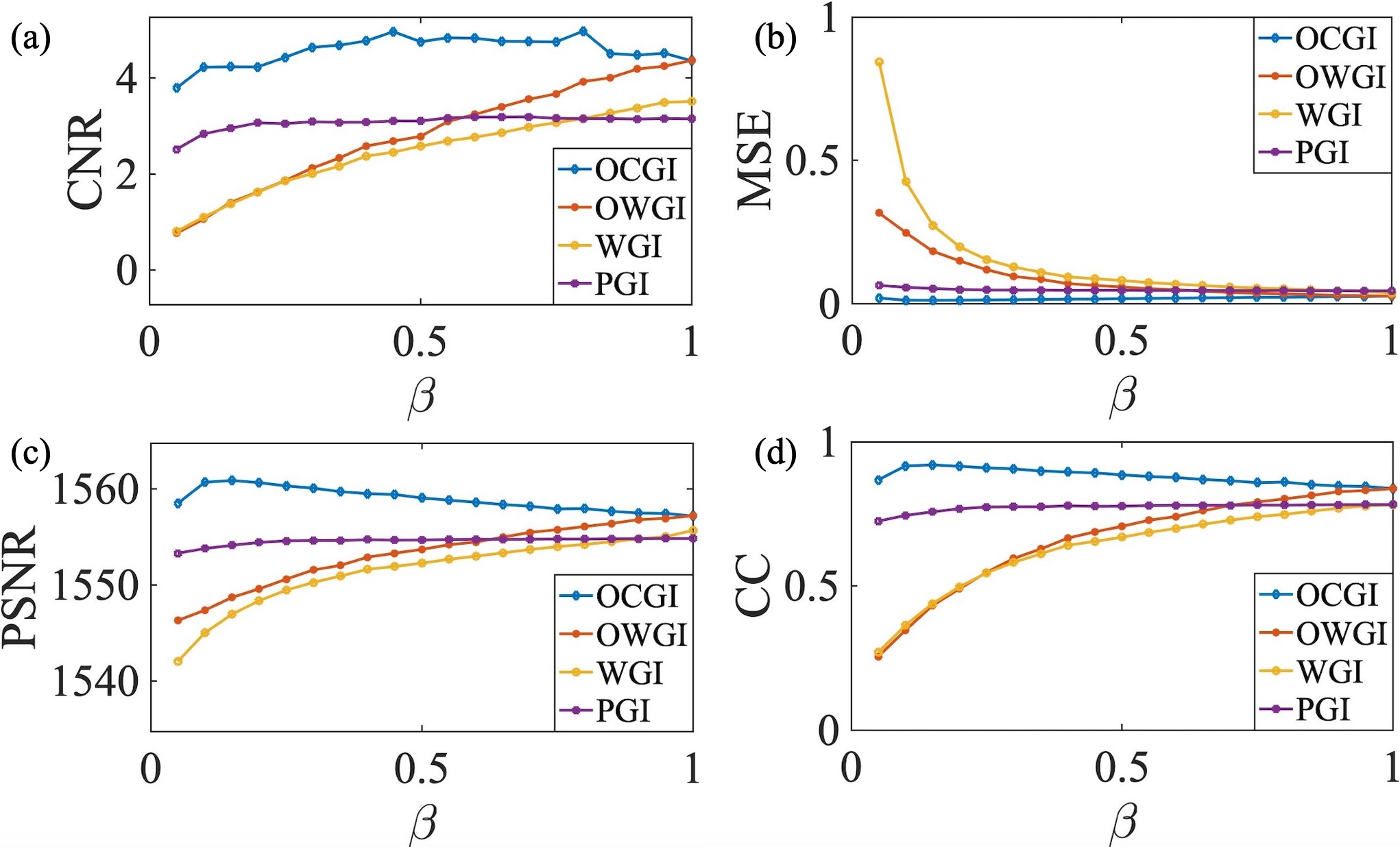}
		\caption{Simulation with added noise. Image qualities via different sampling numbers by CGI with noise at $ 2\%$ signal level. (a) CNR, (b) MSE, (c) PSNR, and (d) CC.}
		\label{fig:simu_noise}
	\end{figure}

	The advantage of OCGI is further demonstrated when we introduce background noise into the system. 
	We run another simulation with a noise level at $ 2\%$ of the signal. The evaluating indicators as functions of $\beta$ are presented in Fig.~\ref{fig:simu_noise}, from which we can see that MSE is about the same as the noise-free case. OCGI still maintains the best in these imaging methods. CNR is dramatically decreased for all methods when $\beta$ is large, as compared to the noise-free case. It should be noted that there are peaks clearly shown in Fig.~\ref{fig:simu_noise}(c) and Fig.~\ref{fig:simu_noise}(d) for OCGI. The PSNR and CC of OCGI reach their highest value when $\beta \sim 0.1$, then slowly decrease and finally reach the same value as that of the OWGI. It suggests that there exist an optimum sampling rate for the noise-free feature during the orthonormalization process of the colored noise pattern. Again, the orthonormalized results are always better than the conventional speckle patterns.

	%\section{Experimental Results}
	\begin{figure}[!ht]
		\centering
		\includegraphics[width=0.95\linewidth]{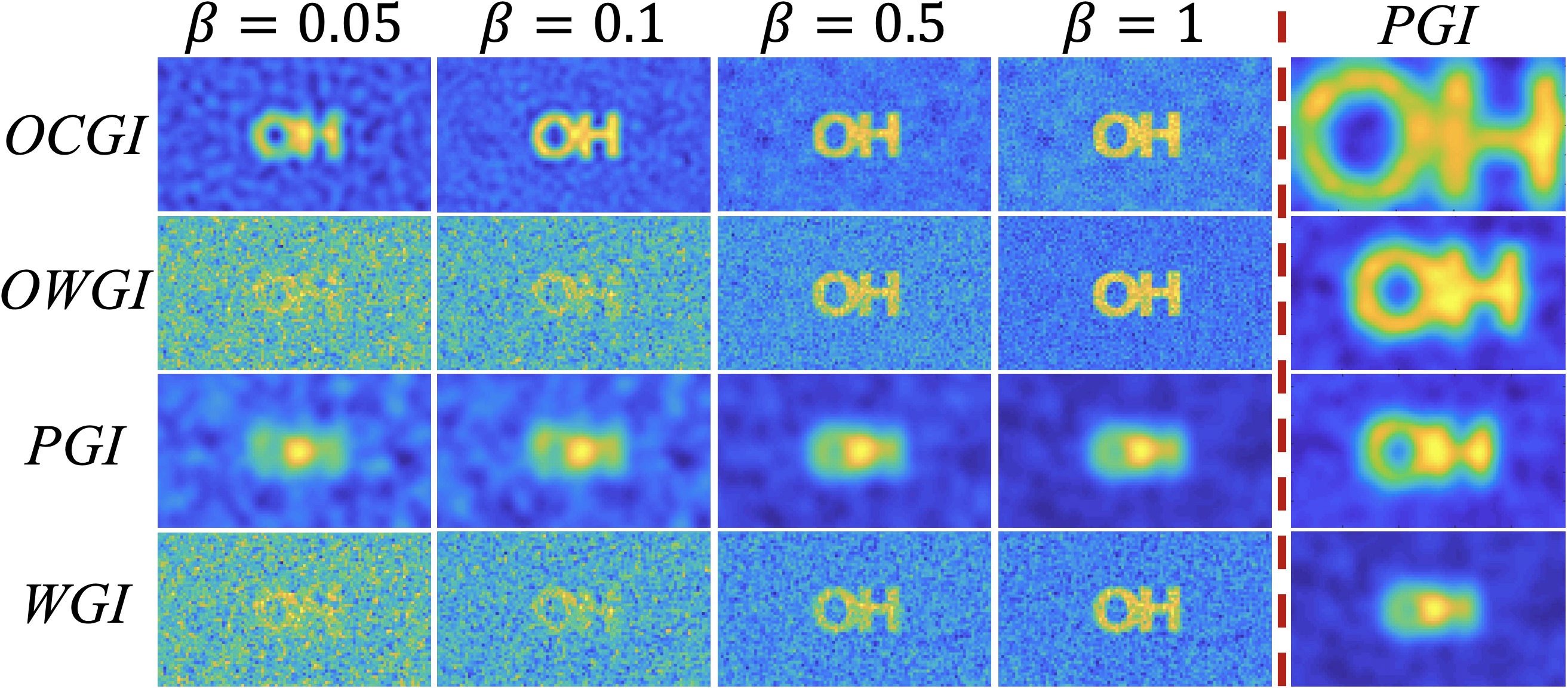}
		\caption{Experimental results. Left-hand side of the red dash line: CGI results via different types of noise pattern with various ${\beta}$. Right-hand side of the red dash line: PGI with different object size at $\beta=1$. the size of the letters, from top to bottom,are: 4, 3, 2, and 1 times of that used for the left side.}
		\label{fig:main_results}
	\end{figure}
We then experimentally test our scheme. In the experiment, we perform measurement on the object `OH'. The noise level is at $\sim 2\%$, which is about the same as the simulation. The main results are shown in Fig.~\ref{fig:main_results}.  ${N_{\mathrm{pixel}}=54\times98}$ is used in our experiment. From Fig.~\ref{fig:main_results}, we see that when $\beta$ is only 0.05, the OCGI already gives an image while all the other methods fail to do so. OCGI, OWGI, and WGI all give clear images at $\beta\sim 0.5$, but the image obtained with OCGI is clearer than OWGI, and both are better than WGI. On the other hand, PGI failed to give a clear image even when $\beta=1$. This is due to the relatively small object size compared with the pixel size. To verify that, we then gradually enlarge the object size 2, 3, and 4 times for PGI at $\beta=1$, as shown in the right-hand side of Fig.~\ref{fig:main_results}. We see that when the object size is large enough, the PGI gives a clear image. We conclude that the image quality in OCGI is better than all the other methods. The PGI method, on the other hand, is limited to the object size and cannot be used for resolution-limit imaging. 
		\begin{figure}[!htp]
		\centering
		\includegraphics[width=0.95\linewidth]{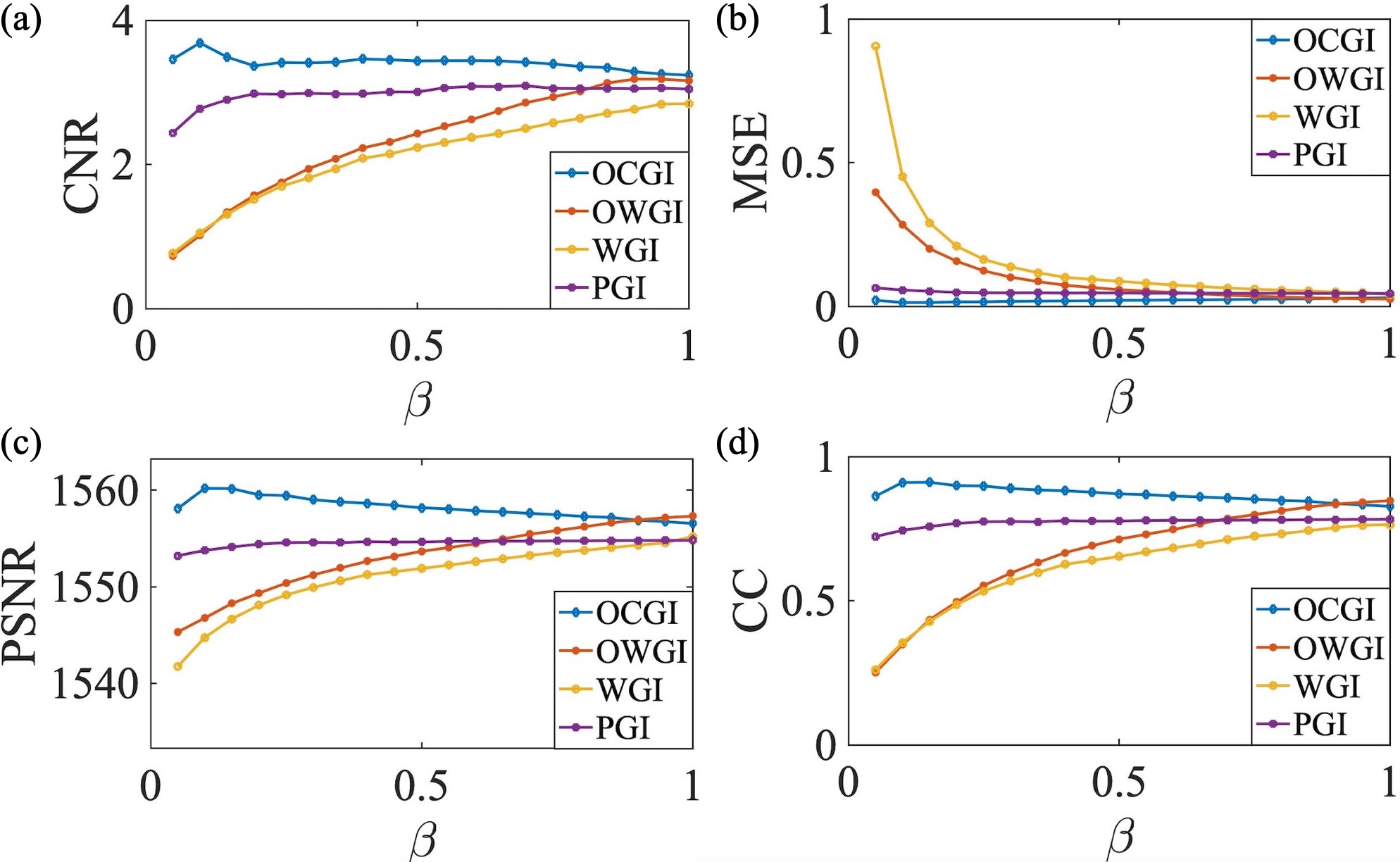}
		\caption{Image qualities via different sampling numbers in the experiment. (a) CNR, (b) MSE, (c) PSNR, and (d) CC.}
		\label{fig:SNR_experiment}
	\end{figure}
	
	To further compare the results, we utilize those four evaluating indicators of image quality again. The results are shown in Fig.~\ref{fig:SNR_experiment}.	We can see that the experimental results and the simulation results are almost exactly matched. We also note here that, as shown in Fig.~\ref{fig:main_results} and Fig.~\ref{fig:SNR_experiment}, some of the indicators suggest the best performance occurs when ${\beta=0.1}$. On the other hand, the results at ${\beta=1}$ seem to have clearer image with sharper edges, which also has the lowest MSE. The reason is when ${\beta=1}$, the cross-correlation disappears, thus no contribution to the area where the object is opaque. Those parameters also give us some indication of the optimal frame rate to choose depending on different experimental goal. 
	
	In conclusion, we have developed a method based on the orthonormalized colored noise pattern in the CGI system to yield image reconstruction results with high quality when the sampling number is small, and with continuously improvement during the further sampling. The major advantage of this scheme is the continuous change of cross-correlation from the orthonormalized colored noise speckle patterns, which overcomes the difficulties faced by the conventional speckle patterns. This method is easy to implement due to its simple setup and rapid image reconstruction. It can reduce the sample rate an order lower than previous orthonormalization methods and it is immune to noise.
	
\textbf{Funding.} Air Force Office of Scientific Research (Award No. FA9550-20-1-0366 DEF), Office of Naval Research (Award No. N00014-20-1-2184), Robert A. Welch Foundation (Grant No. A-1261), National Science Foundation (Grant No. PHY-2013771).
	
\textbf{Disclosures.} The authors declare no conflicts of interest.

\textbf{Data Availability.} Data underlying the results presented in this paper are
not publicly available at this time but may be obtained from the authors upon
reasonable request.
\bibliography{subNyquist_arXiv}

\end{document}